\documentclass{PoS}

\usepackage{graphicx}    

\title{Radio-Quiet AGN and the Transient Radio Sky}

\ShortTitle{Variability in  Radio-Quiet AGN}

\author{\speaker{C.G. Mundell}\thanks{RCUK Academic Fellow}\\
        Astrophysics Research Institute, Liverpool
        John Moores University, U.K.\\
        E-mail: \email{cgm@astro.livjm.ac.uk}}

\author{N.M. Nagar\\
        Universidad de Concepcion, Chile\\
        E-mail: \email{neil@chile}}

\author{P. Ferruit\\
        ESTEC, The Netherlands\\
        E-mail: \email{pierre@ESTEC}}

\abstract{ We present an
  8.4-GHZ radio imaging study of an optically selected sample of early
  type Seyfert galaxies; comparison of images taken at two epochs
  reveals possible variation in the nuclear radio flux density in five
  of them over a seven year period. It is shown that there is a
  possible correlation between the {\em presence} of nuclear radio
  variability and the {\em absence} of hundred parsec-scale radio emission,
  analagous with radio-loud AGN. Our results suggest that all Seyferts
  may exhibit variation in their nuclear radio flux density at 8.4
  GHz, but that variability is more easily recognized in compact
  sources in which emission from the variable nucleus is not diluted
  by unresolved, constant flux density radio jet emission within the
 central ~50 pc.  Taken in combination with other Seyfert properties,
  these results suggest a paradigm of intermittent periods of
  quiescence and nuclear outburst across the Seyfert population and
  demonstrate the importance of investigating the  temporal domain at
  radio wavelengths, which remains completely unexplored for faint
  radio-quiet AGN.  Discovery of intermittent activity and radio flares has important
  implications for the AGN duty cycles and the origin of Ultra-High
  Energy Cosmic Rays. New radio facilities, such as the EVLA/VLBA, eMERLIN, LOFAR and eVLBI on
 the EVN, will revolutionise the study of radio-quiet AGN; in
 particular, the combination of increased sensitivity
 and sampling rate with high-angular resolution and automatic data
 reduction will open up the transient radio sky and bring major future breakthroughs.
}

\FullConference{10th European VLBI Network Symposium and EVN Users Meeting: VLBI and the new generation of radio arrays\\
		September 20-24, 2010\\
		Manchester Uk}

\begin{document}

\section{Introduction}

Variability of nuclear flux density at all wavelengths across the
electromagnetic spectrum has long been recognised as a defining
characteristic of Active Galactic Nuclei (AGN) \cite{sh63}.
The short variability timescales measured at X-ray, optical and UV
wavelengths, ranging from days to years depending on observing
wavelength and intrinsic AGN luminosity, support the standard AGN
paradigm of nuclear emission originating from compact regions powered
by accretion of matter by a central supermassive black hole, rather
than standard stellar processes \cite{sa99,gl03}.

Variability at radio wavelengths is most marked in powerful radio-loud
AGN.  The largest, most rapid variations in nuclear
non-thermal continuum emission are seen in BL Lacs and core-dominated
quasars \cite{li01}. These variations
are explained primarily by Doppler boosting of the nuclear emission by
highly-relativistic jets viewed at small angles to the line of sight
and are consistent with radio-loud Unification Schemes
\cite{ob82}. Flares in the radio light curves might
correspond to ejection of new relativistic components or emergence of
shocks in the underlying flow \cite{bh00}. 

Seyfert nuclei, classified as radio quiet though not radio silent,
have been known for some time to show variability in their continuum
and line emission at X-ray, optical and UV wavelengths
\cite{tu99,wh08}, providing
constraints on nuclear absorbing column variations, photon
reprocessing, structure and dynamics of the Broad Line Region and
 mass of the central object.  However, although ten
times more common than their radio-loud counterparts, Seyferts are 100
to 1000 times weaker at radio wavelengths \cite{go99}; therefore, little is known
about the radio variability of their nuclei as only a small number of
`interesting' Seyferts have been systematically monitored at radio
frequencies, with some
studies motivated by serendipitous discovery of radio flares
\cite{fa00}. 

\begin{figure}
\centering
\includegraphics[height=7cm]{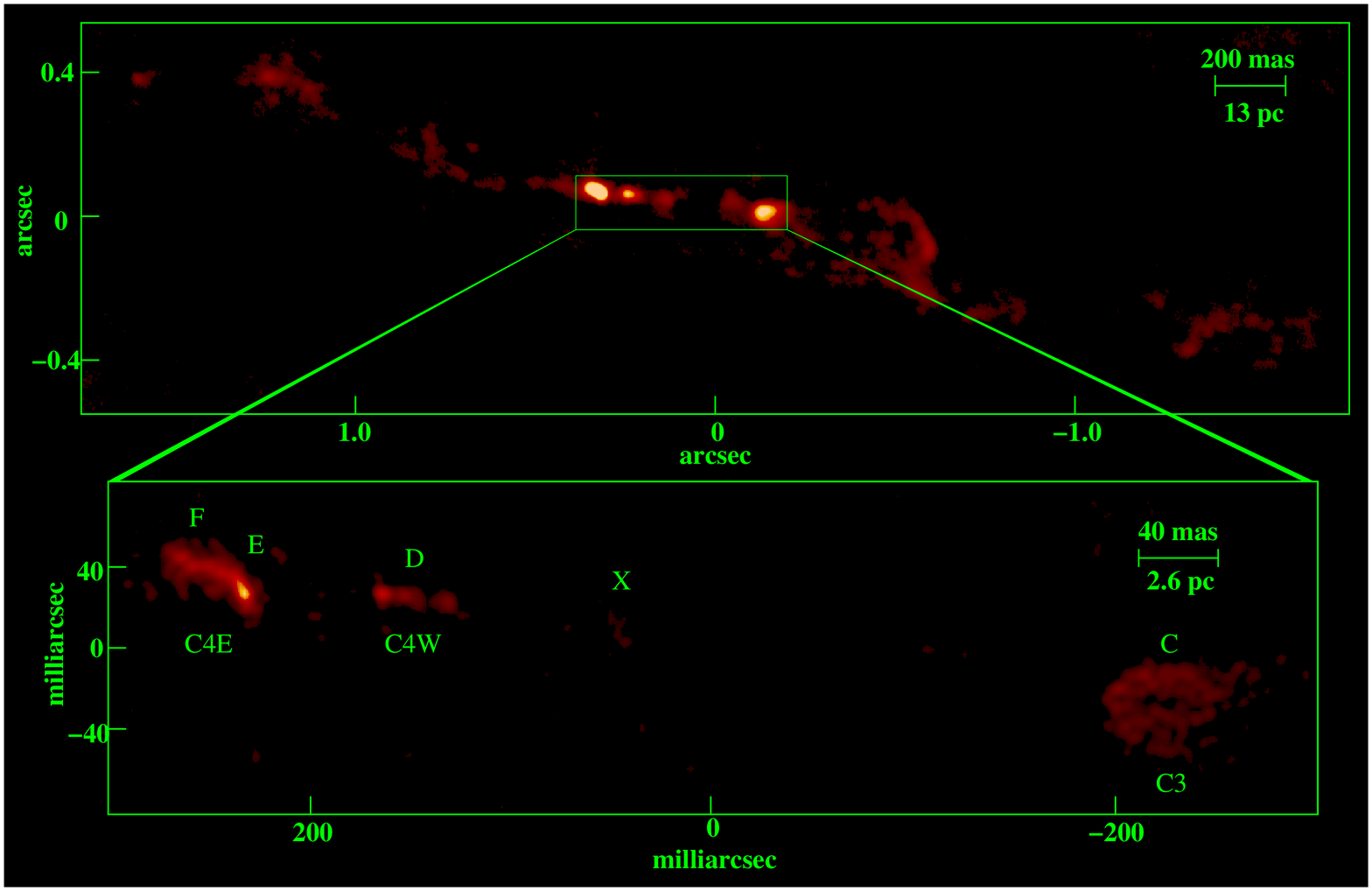}
\caption{VLBA radio images of the highly collimated radio jet in NGC~4151  (from \cite{mn03}.}
\label{fig:N4151}
\end{figure}

Some Seyferts have small radio jets,  $<$1~kpc in extent \cite{uw83,mn03}; with a constant flux density and steep spectral index, this extended radio emission may obscure  fainter flat-spectrum emission associated with the AGN itself unless observed at sub-arcsec angular resolutions \cite{mn95,ch97}.  Identification of compact flat-spectrum radio nuclei with high brightness temperatures using the VLBA, showed unambiguously that Seyfert nuclei are black-hole driven central engines - not compact nuclear starbursts \cite{mn00} - and, coupled with the identification of a highly collimated radio jet in NGC~4151 (Fig. \ref{fig:N4151})  \cite{mn03,ul05}, suggests that radio variability in Seyferts may be common. 

\section{A VLA Search for Nuclear Radio Variability In Seyferts}

We conducted a two-epoch study of 8.4-GHz nuclear
radio emission in a small, but complete sample of 11 nearby
optically-selected Seyferts in which the nuclear radio emission is
imaged at angular resolutions of $\sim$0.2$"$ ($<$50 pc); the
original goal of the study was to obtain high quality radio images to
investigate the impact of Seyfert radio jets on the interstellar
medium by comparing extended radio structures with extended ionised
gas distributions and excitations in the Narrow Line Region as
inferred from images with the Hubble Space Telescope (HST) \cite{fe00}. This comparison is presented in Mundell et al. (2009) \cite{mn09}, including new radio images of MCG~--5-23-16. However, by more reliably separating nuclear
and extended radio components and comparing with previous nuclear flux
density measurements we serendipitously discovered a
fraction of Seyfert nuclei with variable nuclear radio flux densities \cite{mn09}. 

\section{Radio Variability, Seyfert Type and Radio Morphology}

We selected all 12 Seyfert galaxies in early type
hosts with m$_{V}$~$<$~14.5, cz~$<$~3000 km s$^{-1}$ and declination
suitable for VLA observations from Mulchaey's sample of early-type Seyfert galaxies
\cite{mu96}. The optical continuum and ionized gas
properties were studied with the HST \cite{fe00}; all
objects, except MCG~--5-23-16, were observed with the VLA in 1993, at 1.4
and 8.4 GHz, as part of the larger VLA snapshot study of \cite{N99} (hereafter N99) providing a comparison epoch for our observations, which were conducted using the VLA in A configuration at 8.4 GHz on 1999 August 20/21. Full details of data reduction and analysis - including careful and important corrections for tropospheric phase decorrelation - are given in Mundell et al. 2009 \cite{mn09}.
Data presented in N99 were reprocessed using the method presented in \cite{mn09} to allow rigorous comparison of flux density measurements of the weakest sources. The results are as follows:

\begin{itemize}
\item All twelve Seyferts are detected with the VLA; five nuclei show
 a variation in their nuclear flux density between 1992  and 1999 (
NGC~2110, 3081, 5273 and MCG~--6-30-15 declined and NGC~4117
increased). The remaining six - 
Mrk~607, 620, NGC~1386, 3516, 4968 and 7465,
remained constant (one was not observed in 1992). 

\item No correlation is found between the detection of variability and
core luminosity, which lies in the range
8.2~$\times$~10$^{18}$~$<$~L(core)$_{\rm 8.4~GHz}$~$<$~
5.9~$\times$~10$^{21}$~W~Hz$^{-1}$ for this sample. 

\item No correlation is found between Seyfert type and the detection of radio variability.
 Instead, the compactness of the radio emission seems to be a good
 indicator of variability. All variable nuclei, apart from NGC~2110,
 are compact and at best marginally resolved, while nuclei with
 associated extended radio emission have remained constant.

\item NGC~2110 (see Fig. \ref{fig:N2110}) is the only Seyfert in this sample with significant
extended radio emission from well-defined radio jets {\em and} a
variable nucleus, suggesting that the intrinsic nuclear variability is
higher than the observed $\sim$38\%.  The observed nuclear flux
variability indicates significant nuclear structural changes on scales smaller than the VLA
beamsize (i.e. within the central 15 pc), between the
two epochs, possibly due to the appearance and fading of new
components or shocks in the jet. This is consistent
with sub-parsec scale structures identified by \cite{mn00} that consist of a high brightness temperature ($>$~6.0~$\times$~10$^8$~K), synchrotron self-absorbed point source,
extended jet emission 0.1 pc north and south of the nucleus, and a
discrete component $\sim$0.3 pc north of the nucleus, possibly the
first knot in the northern jet. 

\begin{figure}
\centering
\includegraphics[height=8.2cm]{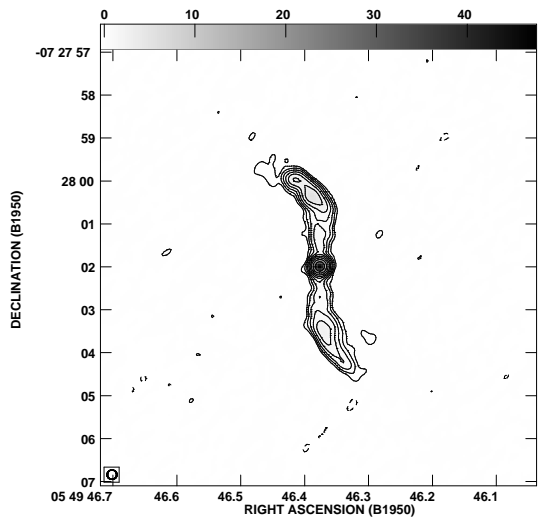}
\vspace{-1cm}
\caption{VLA A-configuration 8.4-GHz radio image of NGC~2110 taken from Mundell et al. (2009).}
\label{fig:N2110}
\end{figure}

\item The serendipitous discovery of variability in $\sim$50\% of our sample, which was not selected on any given radio property or known variability, suggests that radio variability in Seyfert nuclei is a common phenomenon and the measured amplitudes of variation are likely to be lower limits to peak-to-peak variations within this period. This is consistent with studies of III~Zw~2,  which flares approximately every five years\cite{bh03} - a timescale consistent with the variability detected for our sample, given that two-epoch observations are more likely to detect fading sources if the decay timescale is longer than the brightening timescale.

\item All Seyferts might exhibit variations in their nuclear radio flux density at 8.4 GHz, with it being more easily recognised in compact sources in which emission from the
variable nucleus is not mixed with unresolved, constant flux density
radio-jet emission within the $\sim$0.2$''$ VLA beam (
central $\lesssim$~50 pc for V$_{sys}$$<$4000~km~s$^{-1}$).
This is consistent with an increased detection rate of flat spectrum nuclear radio emission in Seyferts imaged at VLBI resolution \cite{mn00,mi04}, further emphasising the importance of isolating nuclear emission from jet emission. Linear resolution limitations for studies of more distant AGN such as radio quiet quasars (RQQ) are more accute. Barvainis et al. \cite{bv05} find few differences between RLQ and RQQ other than an unexpectedly weak dependence of radio variability on nuclear spectral index, with RQQs showing steeper spectral indices than RLQ. If radio quiet quasars have extended, but unresolved steep spectrum emission similar to Seyferts, their nuclear variability and spectral index might have been underestimated, thus increasing their similarity to RL AGN.

\item If flares in radio light curves correspond to
ejection of new relativistic components or emergence of shocks in the
underlying flow, sensitive systematic monitoring of a larger sample,
combined with subarcsecond resolution imaging, may confirm that
Seyferts - as black-hole driven AGN - have the capacity to accelerate
relativistic jets during radio flares despite having radio jets that
are intrinsically non-relativistic during quiescence.
\end{itemize}

\section{Radio Flares and Relativistic Jets}

Flux variability and apparent superluminal motion are defining characteristics of bulk relativistic motion in extragalactic jets. Long-term monitoring of radio-loud AGN \cite{hu92} demonstrated the existence of stochastic, non-random flare events at cm/mm wavelengths, with mean timescales $\sim$2 years for BL Lacs and QSOs \cite{hu92}.  In
 standard jet-shock models,  increases in total flux density are attributed to the passage of shocks on pc scales in collimated, relativistic flows and, in some cases, radio flaring  corresponds to propagation of individual components in VLBI maps \cite{li09}.

In contrast,  debate continues over the nature of ejected radio plasma in Seyferts; the presence of bulk relativistic motion has not yet been demonstrated in Seyfert jets, with existing proper motion studies measuring apparent radio component speeds $V_{app}$~$<$~0.25~c \cite{ul05}. The primary limiting factor in establishing the radio variability characteristics of Seyferts and their radio jets has been the lack of high resolution, long-term monitoring surveys of large samples of Seyferts, whose radio nuclei are faint in quiescence so require careful, phase-referenced observations (e.g. \cite{mn03}).  

Despite the small number of Seyferts studied in detail so far, there is growing evidence that their nuclear activity is characterised by intermittent quiescent phases punctuated by outbursts \cite{wa10}. Serendipitous discovery of radio flares in Mrk~348 and III~Zw~2 - suggest that Seyferts can accelerate their jets to large speeds during an outburst \cite{bh05}.  Most notable was a flare in III~Zw~2  in 1998 that showed a $\sim$30-fold flux increase within 2 years and a highly-inverted radio spectrum peaking at millimeter wavelengths; similar flares occurred at earlier epochs over the last $\sim$25 years at lower frequencies \cite{bh05}. A sudden change in spectral shape in 1999 indicated a structural change in the nucleus -  confirmed by VLBI imaging of the nucleus that revealed a newly ejected radio component expanding at an apparent superluminal speed $V_{app}$~$>$~1.25c \cite{bh00}, in stark contrast to the upper limit of 0.04c measured during quiescence. Discovery of a radio flare in NGC~2110 and the variability of compact Seyfert nuclei in our sample suggests this may be a common phenmenon.

\subsection{Radio Flares in Seyferts as the Origin of Ultra High-Energy Cosmic Rays}

The Auger Collaboration recently reported a significant anisotropy in the arrival directions of Ultra-High-Energy Cosmic Rays (UHECRs) and a possible correlation of events with catalogued local AGN; this is consistent with the theoretical prediction that  energy losses due to photopion production on the cosmic microwave background limit the sources of the highest energy CRs to D$<$100 Mpc. Although particle acceleration in jets and lobes of local radio galaxies offer promising candidates for some of the CR events, many remain controversial and debate centres around whether the coincidence between the relatively large number of local AGN compared with the small number of CR events is mere chance alignment \cite{mo09}. Theoretically, very modest, but {\em intermittent} jet activity in low luminosity AGN, such as Seyferts, could provide the acceleration mechanism for UHECRs \cite{pe09,za09}. In particular, acceleration of UHECRs and the survival of energetic heavy nuclei are possible in the parsec-scale weak jets that are typically observed in radio-quiet AGN. 

Routine, high quality, high angular resolution imaging of a well-selected sample of Seyferts with current and future radio facilities, such as EVLA, eMERLIN, VLBA and LOFAR coupled with multi-messenger follow-ups, would provide valuable input to the debate on the temporal properties of Seyferts, as well as investigating the wider possibility that duty cycles and jet production in local AGN may be associated with UHECRs.

\end{document}